\documentclass{webofc}

\usepackage[varg]{txfonts}   
\usepackage{hyperref}
\usepackage{url}
\usepackage{caption}
\usepackage{subcaption}

\hypersetup{colorlinks=true,citecolor=blue,urlcolor=blue,linkcolor=blue}
\begin{document}
\title{Testing hydrodynamic response to initial-state geometry in Pb+$d^\uparrow$ collisions}
%
%

\author{\firstname{Heikki} \lastname{Mäntysaari}\inst{1,2}
\and \firstname{Björn} \lastname{Schenke}\inst{3}
\and \firstname{Chun} \lastname{Shen}\inst{4}\fnsep\thanks{speaker; \email{chunshen@wayne.edu}}
\and \firstname{Wenbin} \lastname{Zhao}\inst{5,6}
}

\institute{Department of Physics, University of Jyv\"askyl\"a, P.O. Box 35, 40014 University of Jyv\"askyl\"a, Finland
\and Helsinki Institute of Physics, P.O. Box 64, 00014 University of Helsinki, Finland 
\and Physics Department, Brookhaven National Laboratory, Upton, NY 11973, USA 
\and Department of Physics and Astronomy, Wayne State University, Detroit, Michigan 48201, USA
\and Physics Department, University of California, Berkeley, California 94720, USA
\and Nuclear Science Division, Lawrence Berkeley National Laboratory, Berkeley, California 94720, USA
}

\abstract{Deuterons with different polarization states have distinct shapes for their wavefunctions. This offers a unique opportunity to experimentally control the initial-state collision geometry with the polarization of the light-ion targets in relativistic heavy-ion experiments. We study the charged hadron elliptic flow coefficients with respect to the polarization angle of deuterons in Pb + polarized deuteron collisions using a hydrodynamics + hadronic transport model. Hydrodynamic response to initial-state geometry predicts a distinct sign of $v_2$ correlated with the deuteron's polarization states, providing a clean test case for elucidating the collective origin in small collision systems. 
}
\maketitle
\section{Introduction}
\label{intro}

Understanding the emergence of collectivity in small collision systems has been a central topic in the field of high-energy nuclear collisions for more than a decade~\cite{Schenke:2021mxx, Noronha:2024dtq}. Do the observed flow-like correlations in small systems originate from the same collective dynamics of Quark Gluon Plasma (QGP) as in large heavy-ion collisions, or can they be explained by initial-state effects, parton scattering, or other non-hydrodynamic mechanisms? Addressing this question is crucial for advancing our understanding of QCD matter under extreme conditions and for probing the limit of the application of relativistic hydrodynamics~\cite{Chiu:2021muk, Chiu:2025mau}. There have been several attempts to set up controlled collision systems to experimentally identify signatures of hydrodynamic response to initial-state geometry in small systems~\cite{PHENIX:2018lia}. However, the role of event-by-event fluctuations at nucleon and sub-nucleonic levels often makes the interpretation of experimental measurements difficult~\cite{STAR:2023wmd}.

\begin{figure}[h!]
    \centering
    \includegraphics[width=0.49\linewidth]{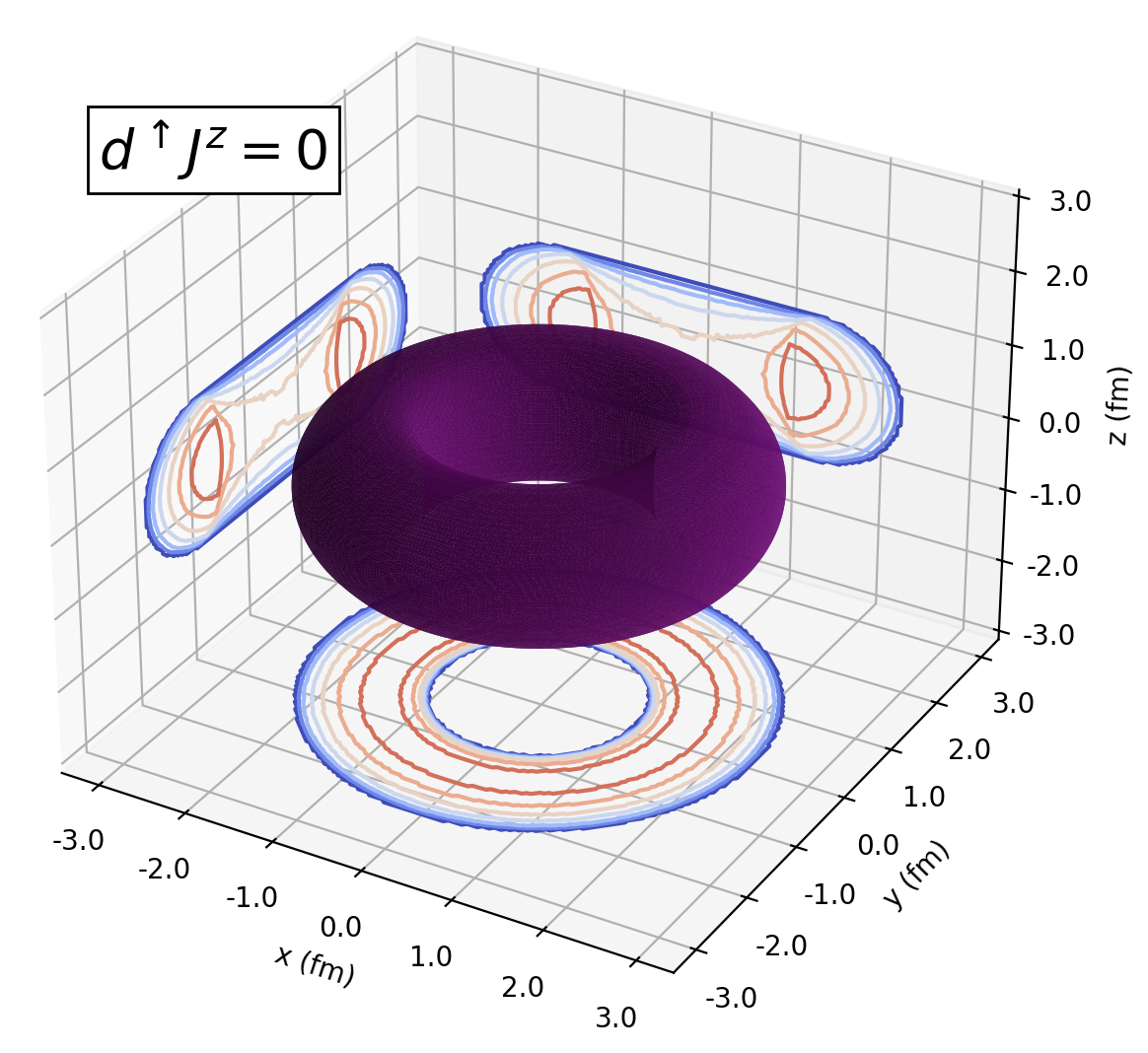}
    \includegraphics[width=0.49\linewidth]{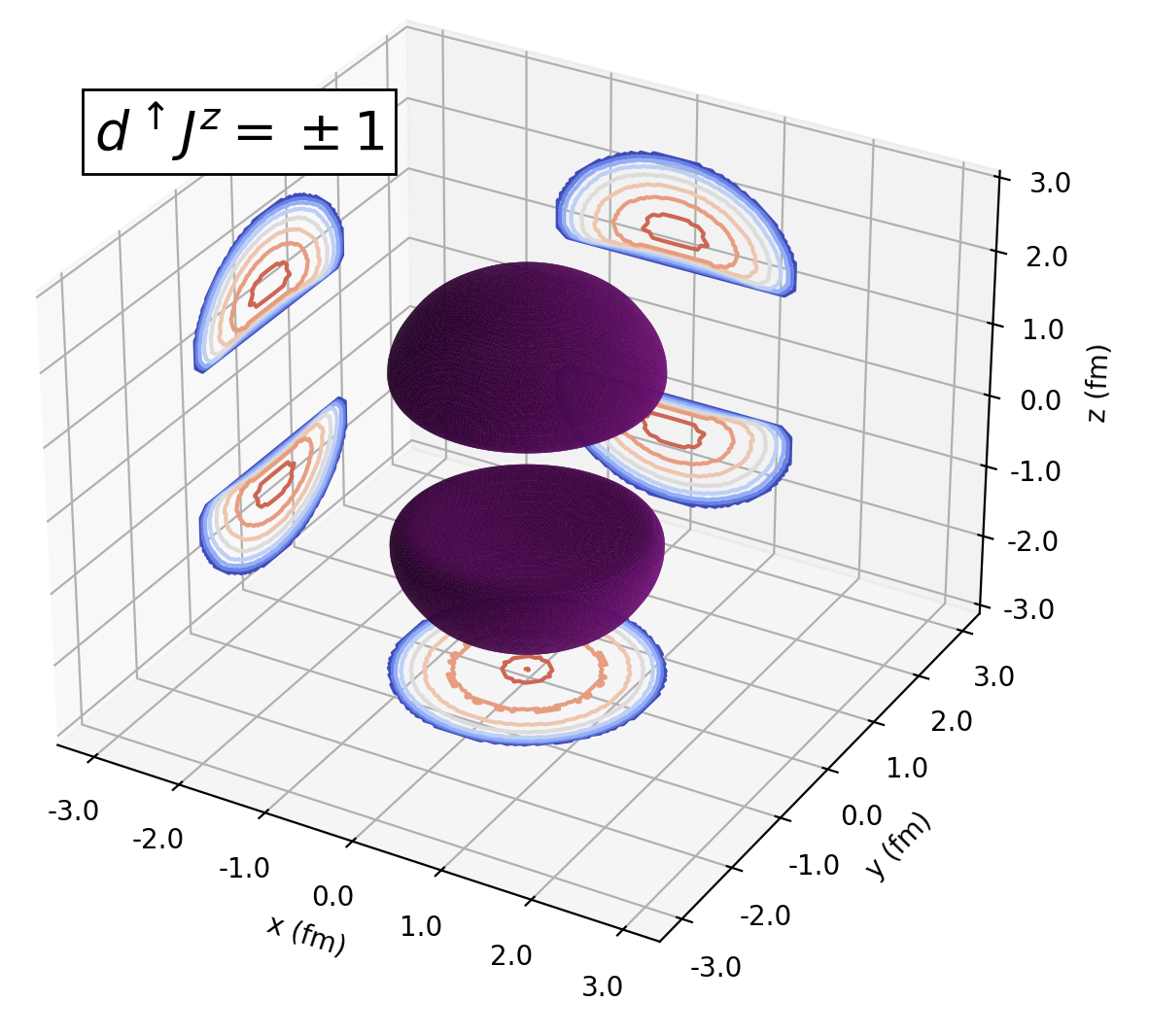}
    \caption{Contours of the polarized deuteron wave function squared for $J^z = 0$ (left) and $J^z = \pm 1$ (right). The parametrization of the deuteron wave function is obtained from Ref.~\cite{Zhaba:2015yxq}.}
    \label{fig:wf}
\end{figure}

Polarization provides a unique tool to control the orientation of light ions before collisions. For example, the deuteron's wave function with different polarization states exhibits distinct spatial shapes as shown in Fig.~\ref{fig:wf}. One way to image the deuteron’s shape in coordinate space is to study the angular dependence of exclusive vector meson production in electron + polarized deuteron collisions~\cite{Mantysaari:2024xmy}. Before such an experiment becomes accessible at the future Electron-Ion Collider, high-energy nuclear collisions with polarized deuterons ($d^\uparrow$) targets offer a new and complementary venue to control the initial-state collision geometry experimentally. Suppose the produced medium from such collisions is strongly coupled. In that case, hydrodynamic response will provide a definite prediction for how the final-state elliptic flow of the collision system should be correlated with the deuteron's polarization state~\cite{Bozek:2018xzy, Broniowski:2019kjo}.  
In this proceeding, we present (3+1)D event-by-event hydrodynamic simulations for Pb + $d^\uparrow$ collisions in the fixed-target mode, which can be experimentally realized with the System for Measuring Overlap With Gas (SMOG) at the Large Hadron Collider beauty (LHCb) experiment.
 
\section{The (3+1)D hybrid framework and results}

Pb+$d^\uparrow$ collisions are effective ways to image the shape of the target light-ions as the relatively large Pb nucleus acts as a sheet to imprint the target deuteron's shape configurations event by event~\cite{Bozek:2018xzy,Broniowski:2019kjo,Giacalone:2024ixe}. Here, we consider colliding Pb nuclei with transversely polarized deuterons, in which the polarization axis is aligned in the transverse plane. With this setup and the illustration in Fig.~\ref{fig:wf}, deuterons with the $J^z = 0$ state will result in an oblate overlapping area with respect to the polarization axis and generate elliptic flow aligned with this axis. On the contrary, deuterons with the $J^z = \pm 1$ states lead to prolate fireballs, and the final-state elliptic flow will point orthogonal to the polarization axis.

In this work, we use the 3D-Glauber + MUSIC + UrQMD hybrid model~\cite{Shen:2017bsr, Shen:2022oyg, Zhao:2022ugy} to simulate the non-trivial longitudinal dynamics and rapidity dependence for the final-state anisotropic flow coefficients in the asymmetric Pb+$d^\uparrow$ collisions. The fixed-target setup in LHCb SMOG would have Pb+$d^\uparrow$ collisions at $\sqrt{s_\mathrm{NN}} = 71$ GeV, which falls within the RHIC Beam Energy Scan program. Therefore, we utilize the highest likelihood model parameters from the recent Bayesian calibration, incorporating the RHIC BES measurements from ~\cite{Jahan:2024wpj, Jahan:2025cbp}. 
To be in line with the LHCb detector acceptances, we compute the collision system's elliptic flow coefficients using charged hadrons with $p_T \in [0.2, 3]$ GeV and pseudo-rapidities $\eta \in [2, 5]$ in the lab frame. 


\begin{figure}[h!]
    \centering
    \includegraphics[width=0.45\linewidth]{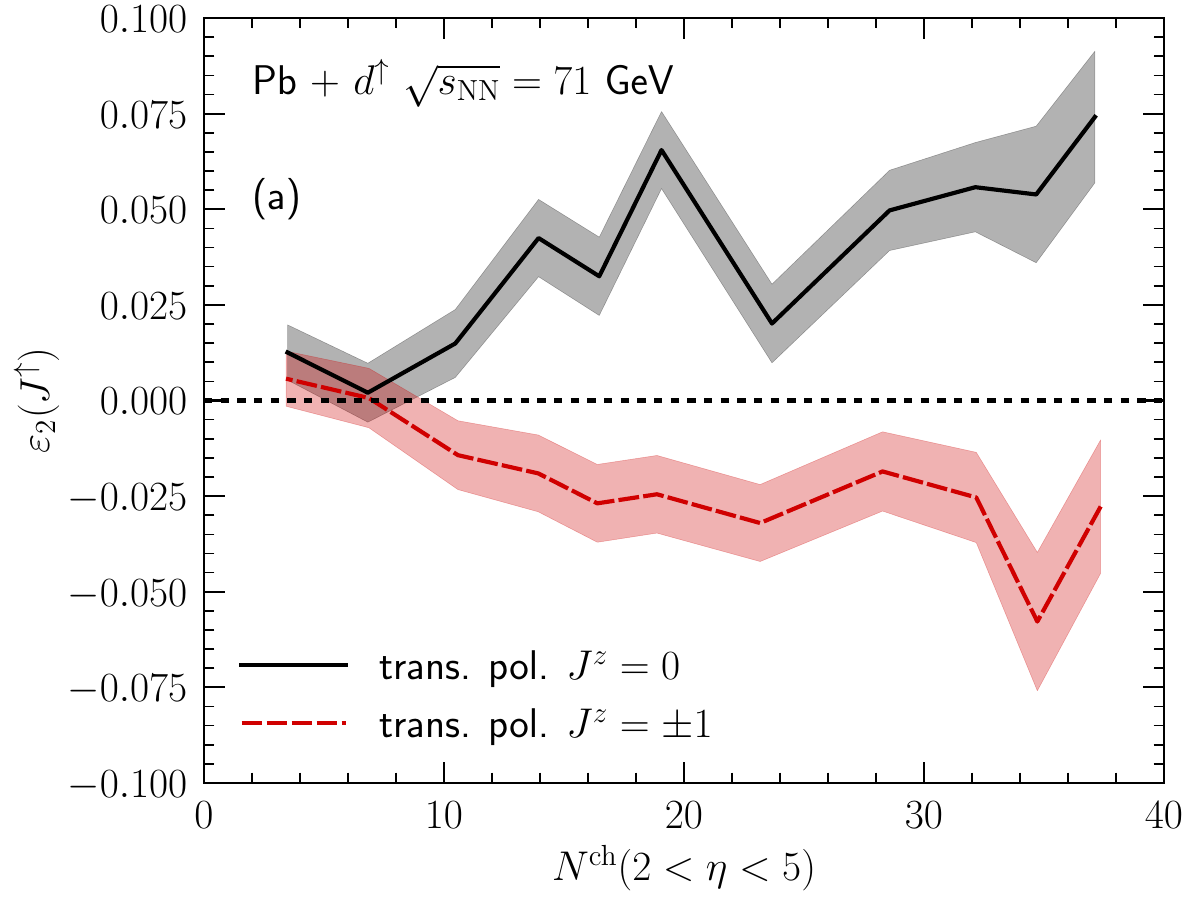}
    \includegraphics[width=0.45\linewidth]{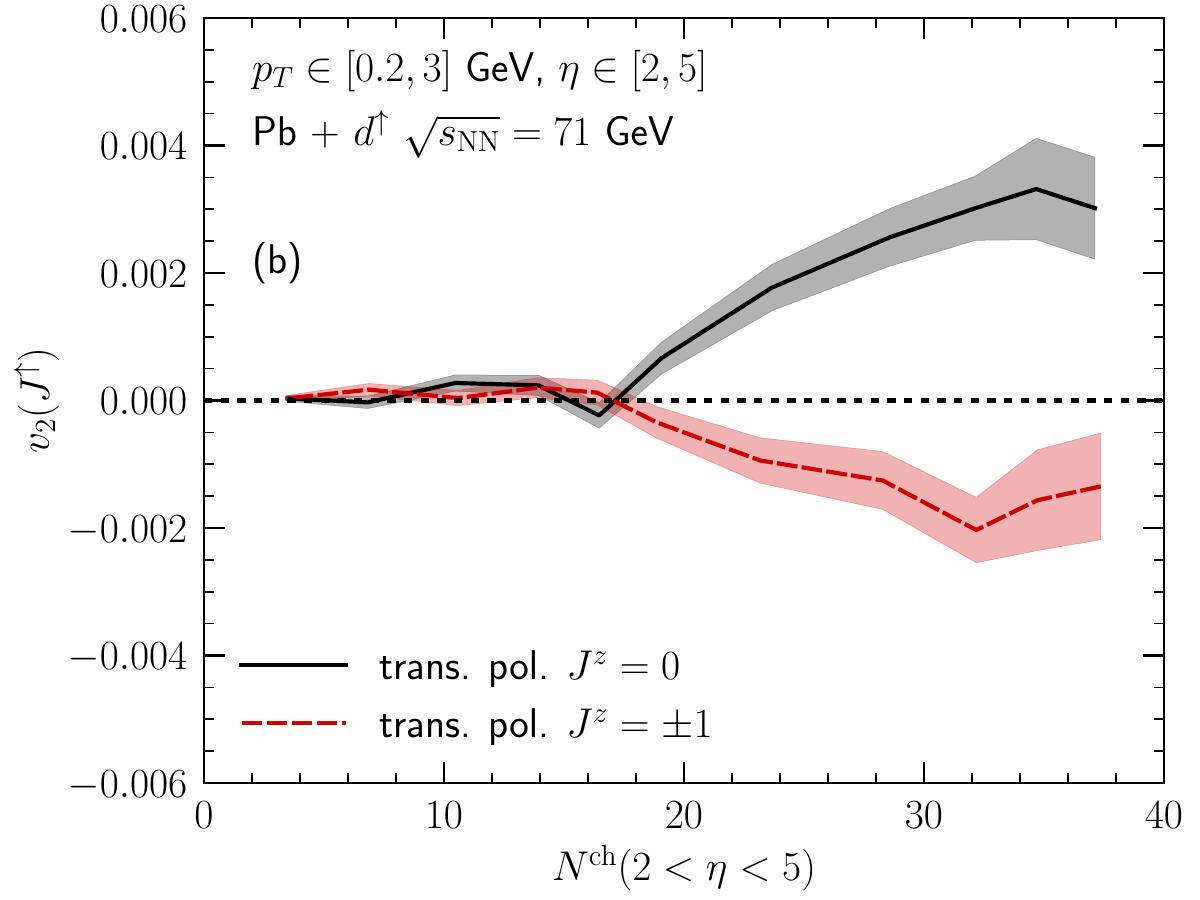}
    \caption{Elliptic flow coefficients measured with respect to the polarization axis for Pb + transversely polarized deuteron collisions at $\sqrt{s_\mathrm{NN}} = 71$ GeV.}
    \label{fig:v2Jz}
\end{figure}

Figure~\ref{fig:v2Jz}a shows the initial state eccentricity $\varepsilon_2$ with respect to the deuteron's polarization axis. For collisions with $N_\mathrm{ch} > 10$, we can clearly observe that the sign of $\varepsilon_2(J^\uparrow)$ coincides with our expectation based on the spatial configurations of deuterons shown in Fig.~\ref{fig:wf}. Figure~\ref{fig:v2Jz}b shows the final-state charged hadron elliptic flow with respect to the deuteron's polarization axis. The magnitudes of $v_2(J^\uparrow)$ are significantly smaller than those of $\varepsilon_2(J^\uparrow)$, indicating the limited hydrodynamic conversion of spatial eccentricity to momentum anisotropy in small systems. Nevertheless, the signs of $v_2(J^\uparrow)$ remain the same as those of $\varepsilon_2(J^\uparrow)$ for different deuteron polarization states. Our results here provide a concrete prediction for the signs and magnitudes of elliptic flow in Pb+$d^\uparrow$ collisions, assuming the system is strongly coupled. We note that the PYTHIA model with string shoving predicts the opposite signs for $v_2$ in these collision systems~\cite{Bierlich:2024lmb}. Therefore, experimental measurements of Pb+$d^\uparrow$ collisions at the LHCb SMOG can provide extremely valuable insights into the origin of collectivity in small systems.

\section{Conclusions}

In this proceeding, we provided a theoretical prediction of the elliptic flow with respect to the deuteron's polarization axis for Pb + transversely polarized $d^\uparrow$ collisions at $\sqrt{s_\mathrm{NN}} = 71$ GeV. Assuming the produced medium is strongly-coupled in these small systems, the hydrodynamic response to the initial-state geometry results in positive $v_2(J^\uparrow)$ for collisions with deuterons with the $J^z = 0$ state, and negative $v_2(J^\uparrow)$ for those collisions with deuterons in the $J^z = \pm 1$ states. Our prediction provides a benchmark result to be compared with upcoming experimental measurements in the LHCspin program with the LHCb SMOG setup, which serves as a concrete test for the hydrodynamic responses in small systems.

\bigskip
\noindent {\it{Acknowledgments.}}
{
This work is supported in part by the U.S. Department of Energy, Office of Science, Office of Nuclear Physics, under DOE Contract No.~DE-SC0012704 (B.P.S) and Award No.~DE-SC0021969 (C.S.), and within the framework of the Saturated Glue (SURGE) Topical Theory Collaboration (B.P.S., W.B.Z.).
C.S. acknowledges a DOE Office of Science Early Career Award.
H.M. is supported by the Research Council of Finland, the Centre of Excellence in Quark Matter, and projects 338263 and 359902, and under the European Research Council (ERC, grant agreements No. ERC-2023-101123801 GlueSatLight and No. ERC-2018-ADG-835105 YoctoLHC).
W.B.Z. is supported by the National Science Foundation (NSF) under grant number ACI-2004571 within the framework of the XSCAPE project of the JETSCAPE collaboration.
This research was done using resources provided by the Open Science Grid (OSG), which is supported by the National Science Foundation awards \#2030508 and \#1836650.
The content of this article does not reflect the official opinion of the European Union and responsibility for the information and views expressed therein lies entirely with the authors.
}

%
\bibliography{refs} 
%
%

\end{document}